\documentstyle[11pt,newpasp,twoside,epsf]{article}
\def\edcomment#1{\iffalse\marginpar{\raggedright\sl#1\/}\else\relax\fi}
\reversemarginpar

\newcommand{\myrule}{\rule[-0.25em]{0pt}{1.3em}}

\begin{document}
\title{X-ray emission and variability of young, nearby stars}
 \author{B. Stelzer, R. Neuh\"auser}
\affil{Max-Planck Institut f\"ur extraterrestrische Physik,
 Giessenbachstrasse 1, D-85741 Garching, Germany}

\begin{abstract}
One of the most prominent properties of young stars is strong and
variable X-ray emission.
The All-Sky Survey performed by {\em ROSAT} and the new high 
sensitivity and high spectral resolution X-ray mission {\em XMM} 
provide complementary information about stellar X-ray sources.
While the spatial completeness of the survey is indispensable for the 
identification of related groups of stars, detailed variability studies and
spectroscopic analysis for individual objects are possible with the
improved effective area, gratings and CCD chips onboard {\em XMM}.
\end{abstract}

\section{Introduction}

X-ray emission has played a major role in the recent identification
of several groups of young, nearby stars. The {\em ROSAT} All-Sky Survey
(RASS) is due to its spatial completeness an important tool for discovering
stellar associations and investigating their X-ray properties. However,
it has comparatively low sensitivity and poor temporal coverage, and is
therefore of limited use for variability studies. The situation is greatly
improved by the new generation of X-ray observatories, e.g. for coronal
X-ray sources {\em XMM} has about an order of magnitude higher 
sensitivity than the RASS and provides much longer exposure times 
allowing continuous monitoring for more than 40\,h.

In this contribution we aim at demonstrating the capabilities of both the RASS
by comparing X-ray luminosity functions (XLF) of different young stellar 
associations, 
and {\em XMM} by discussing one young very nearby star in detail, GJ\,182.

\section{The Tucana Region and Other Nearby Stellar Associations}

The Tucana association (= TucA; Zuckerman \& Webb 2000) is 
one of several recently identified groups of young stars, 
which include the TW\,Hya association (TWA),
the $\eta$ Chamaeleontis cluster ($\eta$\,Cha), 
and the Horologium association (HorA).
Guenther et al. (2001) and more recently Zuckerman et al. (2001) 
suggested that HorA and TucA form a common group.
\begin{figure}[p]
\plotfiddle{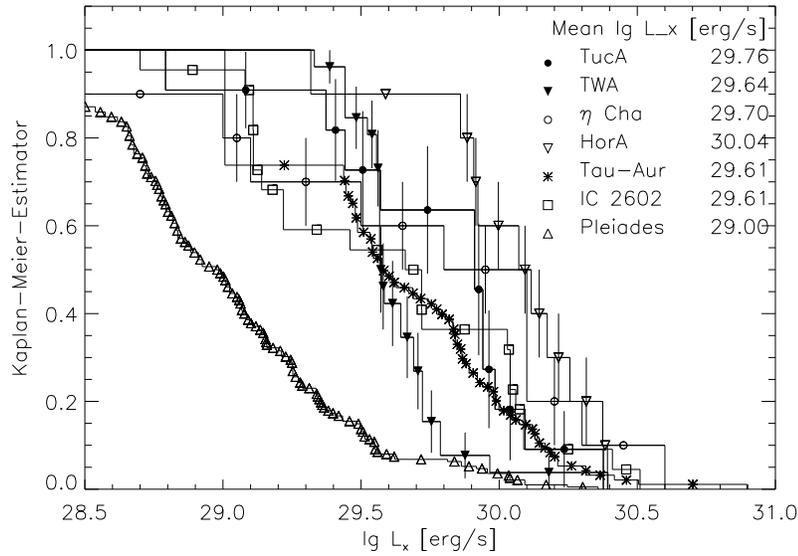}{7cm}{0}{54}{54}{-160}{0}
\caption{XLF of nearby young stellar associations. 
For comparison the XLF of the
Taurus-Auriga star forming region, the IC\,2602 cluster and the Pleiades
are shown. The numbers given in the right of the graph are the mean X-ray
luminosities computed with ASURV (Feigelson \& Nelson 1985) taking account
of upper limits for non-detections.}
\end{figure}
\begin{figure}[p]
\plotfiddle{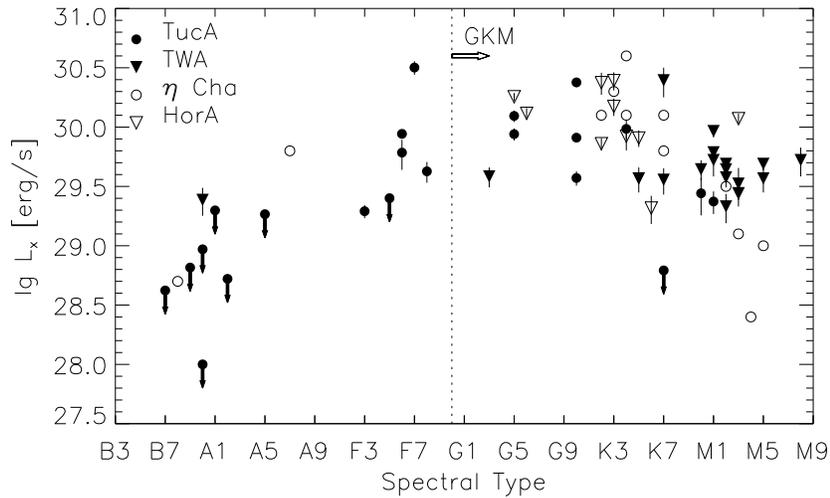}{7cm}{0}{74}{74}{-170}{0}
\caption{X-ray luminosity $L_{\rm x}$ versus spectral type. 
Within the group of stars with significant convection zones (i.e. 
spectral types G and later) $L_{\rm x}$ declines as the stars become
cooler. As expected most A- and B-type stars are undetected in X-rays.}
\end{figure}

We have performed a detailed study of the X-ray emission from the stars 
in TucA making use of the RASS 
(Stelzer \& Neuh\"auser 2000), and have presented the 
first X-ray luminosity functions (XLF) for TucA and TWA. 
Here, we extend our investigation 
to the stars from $\eta$ Cha (Mamajek et al. 1999) 
and HorA (Torres et al. 2000). The XLF for all four groups 
are presented in Fig.~1. The luminosity distributions show significant
overlap, indicating that the average level of X-ray emission is similar
for all groups. The comparison with the XLF from the 
pre-main sequence stars in the Taurus-Auriga region, the young cluster
IC\,2602, and the Pleiades shows that the above 
mentioned four young associations have X-ray luminosities comparable 
to pre-main sequence (PMS) stars 
and are significantly X-ray brighter than the $10^8$\,year-old Pleiades. 
This can be taken as supporting evidence for their youth.
A closer inspection of the XLF reveals the following characteristics:

The HorA sample shows the highest X-ray luminosities. This presumably
demonstrates a selection bias: HorA was found by follow-up observations
of RASS sources, i.e. all members of HorA known so far are X-ray emitters.
X-ray fainter or quiet members in this region are likely to exist, but have
yet to be discovered (see e.g. the contribution by C. 
Torres in this proceedings).
The $\eta$\,Cha group has also been discovered by their X-ray emission,
however, on basis of the more sensitive pointed observations by the {\em
ROSAT} HRI. TucA was identified as a group of stars with common proper
motion and distance, and includes also objects not detected in X-rays, i.e.
the sample is likely to be more complete. 
While several TWA members were also discovered in X-ray selected samples
from the RASS, all the pre-{\em ROSAT} known TTS in TWA were discovered
in optical and IR observations, hence there is no strong X-ray bias in TWA.
The XLF of the TWA stars is steeper than all other distributions. 
This could be due to the very narrow spectral type distribution of the 
known members of this association. 

The relation between X-ray emission and spectral type 
is shown in Fig.~2. Dynamo-driven magnetic activity sets in
around spectral type $\sim$\,F5, as the stars start to have significant 
convective envelopes. Therefore, only stars with spectral type G
and later have been considered in the XLF presented in Fig.~1. 
A substantial number of members
of TucA are A- or B-type stars, and not detected in X-rays in agreement
with theoretical expectations. Among the late-type stars a clear trend
towards declining $L_{\rm x}$ is observed as stars become redder presumably
as a result of smaller (radiating) surface area.

\section{GJ\,182}
 
GJ\,182 is a young field star 
(Gliese 1969) located at a {\em Hipparcos} 
distance of 26.7\,pc, and stands out as particularly Lithium rich 
($W_{\rm Li} \sim 270$\,m\AA; Favata et al. 1997) for its evolutionary age
inferred from standard theory ($\sim 40$\,Myrs; Favata et al. 1998).
It is believed to be one of the most active, nearby stars. Frequent flares
in the U-Band have been reported (de~la~Reza et al. 1981), 
and the $\lg{(L_{\rm x}/L_{\rm bol})}$ ratio
is close to the canonical saturation limit for late-type magnetically 
active stars ($\lg{(L_{\rm x}/L_{\rm bol})} = -3.15$; H\"unsch et al. 1998).

In September 2000 GJ\,182 was observed with {\em XMM}. The
lightcurves in different X-ray energy bands obtained from this 
$\sim$\,17\,ksec 
observation are shown in Fig.~3. Low level variability is observed below
$\sim\,4$\,keV, above this energy the source emits almost no X-rays.
This is also evident from the EPIC-pn CCD spectrum  shown in 
Fig.~4. A 3-temperature model for thermal emission from a hot plasma 
with electron density fixed at $10^{10}\,{\rm cm^{-3}}$ (solid line in
Fig.~4) provides a good description of the data 
($\chi_{\rm red}^2=1.01$, 280 degrees of freedom). The fit 
is acceptable only if the elemental abundances are allowed to vary (see
Table~1 for a summary of the spectral parameters). 
The two lower temperatures are comparable to results from the 
{\em ROSAT} PSPC spectrum during the quiescent state, 
which due to the instruments' restriction to lower energies required 
only two spectral components. During a flare observed by {\em ROSAT} the
higher temperature of the 2-T model increased as a result of coronal
heating, while the lower temperature
remained unaffected. This indicates the presence of a permanent level of
quiescent emission,
as has been observed before on other flaring stars 
(Stelzer \& Hu\'elamo 2000). 

\begin{figure}[p]
\plotfiddle{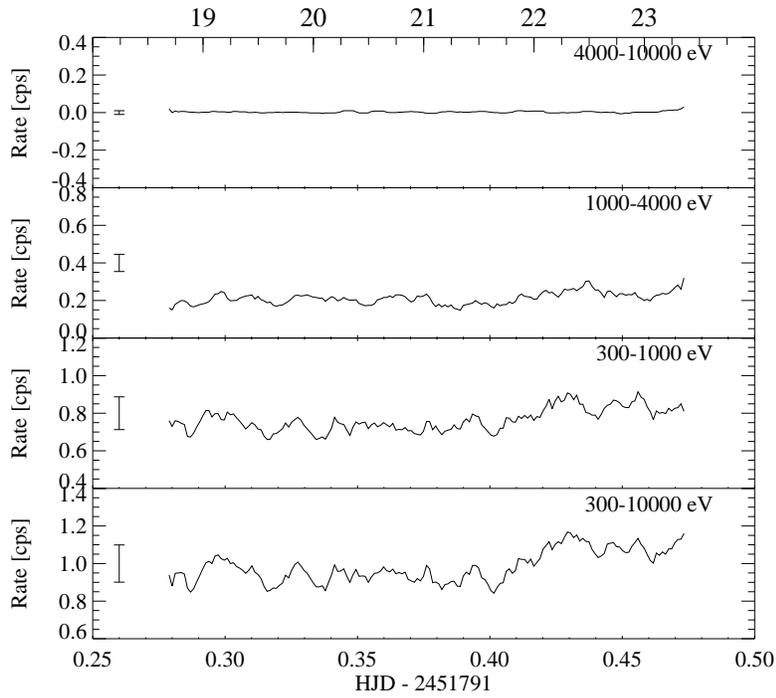}{9.5cm}{0}{50}{50}{-160}{0}
\caption{Smoothed {\em XMM} EPIC-pn lightcurves for GJ\,182 in different energy
bands. The vertical lines at the left side of the plots are the mean errors
in the respective energy band.}
\end{figure}

\begin{figure}[p]
\plotfiddle{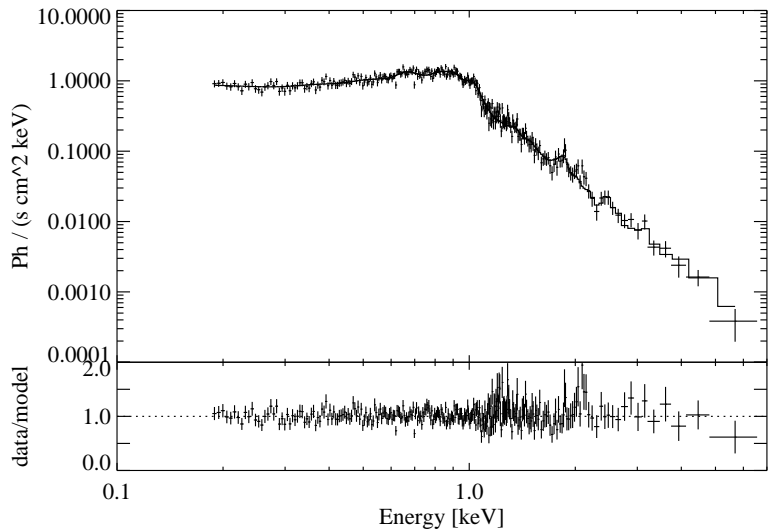}{7cm}{0}{60}{60}{-150}{0}
\caption{{\em XMM} EPIC-pn count rate spectrum of GJ\,182, 
3-temperature best fit model, and residuals.}
\end{figure}

\begin{table}\small
\caption{Spectral parameters for GJ\,182 derived from {\em ROSAT} PSPC and {\em
XMM} EPIC-pn observations. The observation ID is given in brackets.}
\begin{tabular}{l|l|lllll}\hline
\multicolumn{2}{c}{{\em ROSAT} PSPC (200517p/p-1)} \myrule &
\multicolumn{5}{c}{{\em XMM} EPIC-pn (0112880401)} \\ 
\multicolumn{1}{c|}{Flare} \myrule & \multicolumn{1}{c|}{Quiescence} & \multicolumn{5}{c}{Quiescence} \\ \hline
Temp. [keV] \myrule & Temp. [keV] & Temp. [keV] & \multicolumn{4}{c}{Abundances [solar=1.0]} \\ \hline
\myrule $kT_1=0.10^{+0.04}_{-0.03}$ & $kT_1=0.10^{+0.19}_{-0.34}$ & $kT_1=0.21^{+0.05}_{-0.08}$ & O  & $0.48^{+0.31}_{-0.12}$ & Si & $0.48^{+0.21}_{-0.16}$ \\
\myrule $kT_2=0.68^{+0.07}_{-0.05}$ & $kT_2=0.88^{+0.98}_{-0.25}$ & $kT_2=0.60^{+0.03}_{-0.03}$ & Ne & $1.00^{+0.29}_{-0.26}$ &  S  & $0.80^{+0.52}_{-0.41}$ \\
\myrule                            &                             & $kT_3=1.93^{+0.48}_{-0.29}$ & Mg & $0.33^{+0.14}_{-0.14}$ &  Fe & $0.22^{+0.07}_{-0.03}$ \\ \hline
\end{tabular}
\end{table}

\acknowledgments
This work is based on observations obtained with {\em XMM}-Newton, 
an ESA science mission with instruments and contributions directly
funded by ESA Member States and the USA (NASA).
The {\em XMM}-Newton project is supported by Germany's federal government 
(BMBF/DLR), the Max-Planck Society and the Heidenhain-Stiftung.
The {\em ROSAT} project is supported by the Max-Planck Society and BMBF/DLR.

\end{document}